\documentclass[twocolumn,tighten]{aastex62}
\usepackage{amssymb, amsmath,framed}
\usepackage{appendix}

\usepackage{bm}
\expandafter\ifx\csname package@font\endcsname\relax\else
 \expandafter\expandafter
 \expandafter\usepackage
 \expandafter\expandafter
 \expandafter{\csname package@font\endcsname}
\fi
\def\bq{\begin{equation}}
\def\eq{\end{equation}}
\def\bqy{\begin{eqnarray}}
\def\eqy{\end{eqnarray}}

\begin{document}
\title{\Large{Is extraterrestrial life suppressed on subsurface ocean worlds due to the paucity of bioessential elements?}}

\correspondingauthor{Manasvi Lingam}
\email{manasvi.lingam@cfa.harvard.edu}

\author{Manasvi Lingam}
\affiliation{Institute for Theory and Computation, Harvard University, Cambridge, MA 02138, USA}
\affiliation{Harvard-Smithsonian Center for Astrophysics, Cambridge, MA 02138, USA}

\author{Abraham Loeb}
\affiliation{Institute for Theory and Computation, Harvard University, Cambridge, MA 02138, USA}
\affiliation{Harvard-Smithsonian Center for Astrophysics, Cambridge, MA 02138, USA}

\begin{abstract}
The availability of bioessential elements for ``life as we know it'', such as phosphorus (P) or possibly molybdenum (Mo), is expected to restrict the biological productivity of extraterrestrial biospheres. Here, we consider worlds with subsurface oceans and model the dissolved concentrations of bioessential elements. In particular, we focus on the sources and sinks of P (available as phosphates), and find that the average steady-state oceanic concentration of P is likely to be lower than the corresponding value on Earth by a few orders of magnitude, provided that the oceans are alkaline and possess hydrothermal activity. While our result does not eliminate the prospects of life on subsurface worlds like Enceladus, it suggests that the putative biospheres might be oligotrophic, and perhaps harder to detect. Along these lines, potential biospheres in the clouds of Venus may end up being limited by the availability of Mo. We also point out the possibility that stellar spectroscopy can be used to deduce potential constraints on the availability of bioessential elements on planets and moons.\\ 
\end{abstract}

\section{Introduction}
Most studies of habitability, especially from an observational standpoint, rely upon a ``follow the water'' strategy \citep{CB16,WB18}. Whilst this pragmatic approach offers the advantage of simplicity, it is also inherently incomplete since water is only one of the many conditions that are necessary for the chemistry of ``life as we know it''. As a result, some studies have advocated the adoption of the ``follow the energy'' approach as a more comprehensive alternative \citep{HAS07}. Clearly, the availability of energy sources will limit the maximum biomass that can be supported within a given ecosystem. As a result, several studies have sought to quantify the biological potential of habitable worlds by assessing the available energy sources and extrapolating from terrestrial ecosystems \citep{CH01,MF03,MPA08,SI08,PDM17,SDM17,LL18}.

Although the study of available free energy sources is indubitably important, there is one other factor that has not been sufficiently explored: the availability of nutrients. Even when a particular planet (or moon) possesses plentiful free energy, if bioessential elements like phosphorus, nitrogen, and perhaps molybdenum are present in very low concentrations,\footnote{To be more precise, the complete list of bioessential nutrients is not well-defined for ``life-as-we-know-it'', although it seems almost certain that carbon, hydrogen, oxygen, nitrogen, phosphorus and sulfur ought to be placed in this category.} they will set limits on the biological potential of that world. In this paper, we focus on this aspect and study the availability of such elements on worlds with subsurface oceans. Our choice for the latter is dictated by the fact that these worlds are common in our Solar system \citep{Lun17} - most notably Europa and Enceladus - and there are many missions that are either in development, for e.g. Europa Clipper,\footnote{\url{https://www.jpl.nasa.gov/missions/europa-clipper/}} or under review. In addition, several recent observational breakthroughs have revealed that these worlds possess many of the requisite ingredients associated with habitability including liquid water, energy sources and complex organics \citep{NP16,WG17,SS17,PKA18}.

The outline of the paper is as follows. In Sec. \ref{SecNut}, we provide a brief introduction to the candidates for limiting nutrients on subsurface ocean worlds. We introduce a simple mathematical model in Sec. \ref{SecSSC} for computing the steady-state concentrations on these worlds by modelling the sources and sinks, with a particular emphasis on phosphorus as the limiting nutrient. In Sec. \ref{SecSearch}, we outline how the inventory of bioessential elements on exoplanets could be estimated, and briefly comment on the habitability of the Venusian atmosphere. We conclude with a summary of our results in Sec. \ref{SecConc}.

\section{Ocean worlds and nutrients}\label{SecNut}
We start by describing how ocean productivity is governed by the availability of limiting nutrients such as phosphorus and nitrogen on Earth.

\subsection{Limitations on ocean productivity}
There has been a great deal of debate as to what constitutes the limiting nutrient (LN) insofar the total ocean productivity is concerned. This is not an easy question to answer since nutrient limitation can be manifested in different ways, and across varied timescales. The scarcity of nutrients can affect the growth rates of individual cells and the theoretical upper bound on the biomass, corresponding to the Blackman and Liebig limitations. In recent times, the significance of nutrient co-limitation has also been emphasized, wherein two or more elements limit ocean productivity. Moreover, different elements may serve as the limiting factors depending on the timescales under consideration - this prompted \citet{Tyrr99} to distinguish between the proximate and ultimate limiting nutrients, with the latter governing ocean productivity over long timescales.

On Earth, the net primary productivity (NPP) is proportional to the availability of nutrients via the factor $\gamma$ that quantifies the effect of nutrient availability on the maximum growth rate \citep{SG06}. It is common to model $\gamma$ using the Monod equation:
\begin{equation}
    \gamma = \frac{\phi_e}{\mathcal{K}_e + \phi_e},
\end{equation}
where $\phi_e$ is the concentration of the limiting element `$e$' and $\mathcal{K}_e$ is the corresponding Monod constant. This function is well-justified on observational grounds, as seen from Figure 4.2.6 of \citet{SG06}. An immediate consequence of the above discussion is that $\mathrm{NPP} \rightarrow 0$ when $\phi_e \rightarrow 0$ in the context of this simple model. Hence, in this limit, any biospheres (if present) may be highly oligotrophic, and would therefore have a relatively low likelihood of being detectable.

\subsection{Which element limits total productivity?}
Bearing in mind the complexity of nutrient limitation, the three most prominent candidates for the limiting element are nitrogen (N), phosphorus (P) and iron (Fe). Traditionally, P (in the form of phosphates) has been considered to be the limiting element by geochemists \citep{PGH17} whereas N (in the form of nitrates) has been regarded as the limiter by biologists.\footnote{In what follows, we will use P as the shorthand notation for phosphates and N for nitrates wherever appropriate.} Support for N comes from the fact that nitrate is depleted slightly prior to phosphate, and the addition of nitrate stimulates growth in many nutrient-limited environments while the addition of phosphate does not produce an equivalent effect. 

Those who argue in favor of P as the limiting element have pointed to the absence of nitrogen fixation and denitrification analogues for phosphates. Hence, the availability of P is wholly contingent on its delivery from external sources, and has been argued to be the ultimate limiting nutrient defined earlier \citep{Tyrr99}. On Earth, recent evidence suggests that the rise in oxygen levels and the diversification of animals coincided with a fundamental shift in the phosphorus cycle in the late Proterozoic era around $800$ to $635$ Ma \citep{PRB10,RPG17,CLM18}. Prior to this period, there is widespread evidence indicating that the availability of P was much lower than today, consequently suppressing primary productivity \citep{KS17}. If the emergence of animals was indeed linked with the P cycle, the availability of nutrients (especially phosphorus) might be directly tied to animal evolution on Earth \citep{Kno17}, as well as on other exoplanets. For the above reasons, we shall focus mostly on the sources and sinks of P henceforth and explore the ensuing implications.

A strong case can also be made for Fe as the limiting nutrient \citep{BE10,TBB17}. It forms a central component of proteins used in respiration and photosynthesis, and enzymes involved in fixing nitrogen and using nitrates. Consequently, lower levels of Fe could, in turn, lead to reduced growth rates and uptake of other nutrients (such as nitrogen) and thereby set limits on the total productivity \citep{AK02}. However, the availability of Fe is contingent on its solubility in water as noted in Sec. \ref{SSecON}. Apart from Fe, other trace metals such as molybdenum (Mo), manganese (Mn) and cobalt (Co) may also serve as the limiting nutrients during certain epochs \citep{Anbar,RLP16}, but the number of observational and theoretical studies involving these elements is limited when compared to N, P and Fe.

\section{Sources and sinks for the limiting elements}\label{SecSSC}
We will work with the following idealized model to describe the evolution of nutrient availability over time:
\begin{equation}
    \frac{d \mathcal{C}_e}{dt} = \mathcal{S}_e - \mathcal{L}_e \mathcal{C}_e,
\end{equation}
where $\mathcal{C}_e$ (in mol) is the total amount of the nutrient in the ocean, whereas $\mathcal{S}_e$ represents the net influx (sources) of the nutrient (in mol/yr) and $\mathcal{L}_e$ denotes the net outflux (sinks) of the nutrient (in yr$^{-1}$). The subscript `$e$' refers to the limiting element under consideration. The steady-state value of $\mathcal{C}_e$ that will be finally attained is found from setting the RHS to zero, which leads us to
\begin{equation}\label{Cequ}
    \mathcal{C}_e = \frac{\mathcal{S}_e}{\mathcal{L}_e}
\end{equation}
It is therefore instructive to consider the major sources and sinks on Earth, and ask whether they would be present on worlds with subsurface oceans.\footnote{We have addressed this issue briefly in \citet{LL18}, but elements other than P as well as abiotic sources were not considered therein.} Broadly speaking, the external inputs of nutrients to the oceans arise from three different sources on Earth: fluvial (from rivers), atmospheric and glacial. Of the three, the first two sources are much more important than the third (except for P) on Earth, as seen from Table 1 of \citet{MMA13}. However, neither riverine nor atmospheric inputs are directly accessible to worlds with subsurface oceans. Although each nutrient will be depleted through multiple (and different) channels, the burial of organic sediments constitutes a common sink for all the elements \citep{SB13}.

\subsection{Phosphorus: sources and sinks}\label{SSecPho}
We continue our analysis by considering the sources and sinks of P. An important abiotic sink for P is hydrothermal activity \citep{FBL82}. On Earth, the major hydrothermal processes responsible for P removal include low-temperature oceanic crust alteration via ridge-flank hydrothermal systems and scavenging by Fe oxyhydroxide particles in hydrothermal plumes \citep{Del98}. As we know that hydrothermal activity is currently existent on Enceladus \citep{HP15,WG17}, and possibly on Europa, we will restrict our treatment to subsurface ocean worlds where these mechanisms are expected to be operational. A major advantage accorded to this class of icy worlds is the existence of hydrothermal vents because they are widely regarded as promising sites for the origin of life \citep{MBK08,RBB14,SHW16}.

In order to compute $\mathcal{L}_P$ for hydrothermal activity, it is instructive to demonstrate how this estimate is carried out on Earth. On Earth, most of the depletion occurs via low-temperature ridge-flank hydrothermal interactions. The corresponding heat flow in this region is $Q_{HV} \sim 8 \times 10^{12}$ W \citep{SS94}, and the water is raised by $\Delta T \sim 10$ K \citep{WFM96}. Thus, assuming that all the heat is used up for raising the temperature of the water flowing through the hydrothermal systems, we find that the total circulation $F_h$ is
\begin{equation}\label{HydCirc}
    F_h = \frac{Q_{HV}}{c_w \Delta T} \sim 6.3 \times 10^{15}\, \mathrm{kg/yr}
\end{equation}
where $c_w \approx 4 \times 10^3$ J kg$^{-1}$ K$^{-1}$ denotes the heat capacity of water at this temperature. In the idealized limit, all of the P present in the seawater will be precipitated and removed. On Earth, the seawater concentration of P, depleted through hydrothermal vent activity, is assumed to be around $2\,\mu$M \citep{WFM96},\footnote{Note that $1 \mu$M is one of the common units for molality, and is defined as $10^{-6}$ moles of solute per one kilogram of solvent (water in our case); similarly, by definition, note that $1\,\mathrm{mM} = 10^{3}\,\mu$M and $1\,\mathrm{nM} = 10^{-3}\,\mu$M.} and multiplying this with $F_h$ yields a total P removal of $1.2 \times 10^{11}$ mol/yr, which differs from the more recent empirical estimate only by a factor of $2.5$ \citep{WM03}, and we will therefore adopt the same formalism for other worlds. Lastly, for the purpose of our order-of-magnitude estimate, we can model the average concentration as $\phi_P \approx \mathcal{C}_P/M_\mathrm{oc}$, where $M_\mathrm{oc}$ is the mass of the ocean. We make use of the fact that $F_h \cdot \phi_P \equiv \mathcal{L}_P \cdot \mathcal{C}_P$, because the RHS and LHS are merely different representations of the total loss rate (in mol/yr), thereby yielding
\begin{equation}\label{AbSink}
    \mathcal{L}_P \approx \frac{F_h}{M_\mathrm{oc}}.
\end{equation}
Thus, we are now in a position to compute $\mathcal{L}_P$ for an arbitrary subsurface ocean world. 

We make use of the fact that $F_h \propto Q_{HV}$ and $M_\mathrm{oc} \propto R^2 \mathcal{H}$, with $R$ and $\mathcal{H}$ denoting the radius and average ocean depth of the subsurface world. We are still confronted with $Q_{HV}$, which still lacks a straightforward scaling in terms of basic parameters. Although this value is bound to depend on the specific characteristics of the ocean world under consideration, it is ultimately derived from radiogenic heating. Since the latter is conventionally assumed to be linearly proportional to the mass $M$ \citep{VOC09}, we will use $Q_{HV} \propto M \propto R^{3.3}$; the last scaling follows from the mass-radius relationship for icy worlds smaller than the Earth \citep{SGM07}. By substituting the above scalings into (\ref{AbSink}), we obtain
\begin{equation}\label{LPH}
    \mathcal{L}_P \sim 1.7 \times 10^{-5}\,\mathrm{yr}^{-1}\,\left(\frac{R}{R_\oplus}\right)^{1.3}\left(\frac{\mathcal{H}}{1\,\mathrm{km}}\right)^{-1}
\end{equation}

Next, we turn our attention to abiotic source(s) for P. The primary source is expected to be submarine weathering (of felsic rock), but there is a crucial difference: on Earth, continental weathering is via rain water with a pH of $5.6$ whereas submarine weathering occurs through sea water, whose pH is assumed to be approximately $8.0$. The dissolution rate per unit area ($\Gamma$) of phosphate-producing minerals can be estimated from
\begin{equation}\label{Rate}
    \log \Gamma = \log k_{H_+} - n_{H_+} \mathrm{pH},
\end{equation}
where $\log k_{H_+} \approx -4.6$ is the logarithm of the intrinsic rate constant and $n_{H_+} \approx 0.9$ is the reaction order for combination of the minerals chlorapatite, merrillite, whitlockite and fluorapatite \citep{AHF13}. Denoting the dissolution rates on Earth and subsurface worlds by $\Gamma_E$ and $\Gamma_{SS}$ respectively, and introducing $\Delta = \mathrm{pH}_{SS} - \mathrm{pH}_E$ (where it must be recalled that $\mathrm{pH}_E = 5.6$) and $\delta = \Gamma_{SS}/\Gamma_E$, we obtain
\begin{equation}\label{defdel}
    \log \delta = -0.9 \Delta,
\end{equation}
and if we use $\Delta \approx 2.4$ based on the above considerations, we find that $\delta \approx 7 \times 10^{-3}$. Estimating the dissolved preanthropogenic P input from continental weathering is difficult, and we adopt the value $\sim 1.3 \times 10^{10}$ mol/yr for the Earth that is consistent with Fig. 2 of \citet{PM07}; see also \citet{BN00} and \citet{SMB10} for a discussion of this issue. Assuming that the area of weathered regions is proportional to the total area and using the above data, the net delivery of P to the ocean can be expressed as
\begin{equation} \label{NPA}
\mathcal{S}_\mathrm{P} \sim \,1.3 \times 10^{8}\,\mathrm{mol/yr}\,\left(\frac{\delta}{0.01}\right)\left(\frac{R}{R_\oplus}\right)^2,
\end{equation}
and we have normalized $\delta$ by its characteristic value. 

We are now in a position to use (\ref{LPH}) and (\ref{NPA}) in order to calculate (\ref{Cequ}). By doing so, we end up with
\begin{equation}
    \mathcal{C}_P \sim 7.6 \times 10^{12}\,\mathrm{mol}\,\left(\frac{\delta}{0.01}\right)\left(\frac{R}{R_\oplus}\right)^{0.7}\left(\frac{\mathcal{H}}{1\,\mathrm{km}}\right).
\end{equation}
As a consistency check, choosing $\delta = 1$ and $\mathcal{H} \approx 3.7$ km for the Earth leads us to $\mathcal{C}_P \sim 2.8 \times 10^{15}$ mol, which is nearly equal to the empirically estimated value \citep{BN00,PM07}. Instead of the total amount of P in the ocean, we are more interested in the average concentration $\phi_P \approx \mathcal{C}_P/M_\mathrm{oc}$ introduced earlier. Upon solving for this variable, we find
\begin{equation}\label{phip}
    \phi_P \sim 20\,\mathrm{nM}\,\left(\frac{\delta}{0.01}\right)\left(\frac{R}{R_\oplus}\right)^{-1.3}.
\end{equation}
An interesting feature of this formula is that it does not depend on the average ocean depth, and is a monotonically decreasing function of the radius. At first glimpse, $\phi_P$ appears to be around two orders of magnitude lower than the typical P concentrations observed in Earth's ocean, but it must be recognized that it exhibits a very strong (i.e. exponential) dependence on the pH through $\delta$. To illustrate this point further, let us evaluate $\phi_P$ for Europa and Enceladus. 

In the case of Europa, it has been proposed that oxidants formed on the surface via radiolysis are delivered to the subsurface ocean, where they react with sulphides and give rise to a highly acidic ocean with a pH of $2.6$ \citep{PG12}. In this case, we obtain $\delta \approx 5 \times 10^2$, thereby leading us to $\phi_P \sim 6.1$ mM. Hence, when viewed solely from the standpoint of P availability, Europa does not seem to pose any issues for habitability. However, there are other detrimental effects arising from a highly acidic ocean that have been discussed in \citet{PG12}.

The situation is rendered very different when we consider Enceladus. A theoretical model concluded that the ocean comprised of a Na-Cl-CO$_3$ solution with a pH of $\sim 11$-$12$ \citep{GBW15}. The alkanine (high pH) nature of the ocean was argued to stem from the serpentinization of chondritic rock. If we choose a pH of $11.5$, we find that $\delta \approx 4.9 \times 10^{-6}$, and using (\ref{phip}) leads us to $\phi_P \sim 0.65$ nM. The concentrations of total dissolved phosphorus (TDP) - commonly defined as the sum of dissolved inorganic phosphorus (SRP) and dissolved organic phosphorus (DOP) - in many oligotrophic biospheres (e.g. the North Pacific Gyre) are typically $\sim 10$-$100$ nM as seen from Table 5.2 of \citet{KB14}. However, there exist some cases where oligotrophes are expected to be functional even at sub-nM concentrations of dissolved P \citep{HTS00,MT02,KB14}. Hence, there appear to be sufficient grounds to suppose that life-as-we-know-it may exist on Enceladus even at the very low concentrations predicted by our model.

A case could be made for testing this hypothesis by sampling the Enceladus plume or ocean. While H$_2$ \citep{WG17} and complex organic molecules above 200 daltons \citep{PKA18}, including potentially nitrogen-bearing species, have been detected in the Enceladus plume by the \emph{Cassini} spacecraft, phosphorus has not yet been found. Recently, the ROSINA (Rosetta Orbiter Spectrometer for Ion and Neutral Analysis) mass spectrometer was able to detect phosphorus (presumably in the form of PH$_3$) in the coma of a comet (67P/Churyumov-Gerasimenko) for the first time \citep{AB16}. While this does open up the possibility of carrying out similar observations of the Enceladus plume, it should be noted that the predicted abundance of PH$_3$ relative to water was quite high ($\sim 10^{-3}$) in the coma. Hence, given that our predictions yield a much lower steady-state concentration of P in the Enceladus ocean, the mass spectrometer aboard a future mission will need to be highly sensitive. In contrast, if \emph{in situ} analysis of the ocean can be carried out, even sub-nM concentrations are readily measurable by systems with liquid waveguide capillary cells and miniaturized spectrophotometers \citep{ZC02,PRS08}.

In \citet{LL18}, the possibility of a P source analogous to glacial weathering was considered, but its likelihood was deemed to be low for generic subsurface ocean worlds. Next, we turn our attention to an important \emph{biotic} sink: the burial of organic sediments. Estimating this value is not an easy task, since the burial rate is subject to significantly spatio-temporal variability and dependent on a number of environmental factors. However, one can envision two general regimes:\\

{\bf Case I:} If we assume that the burial rate is proportional to the biomass present in the oceans, the latter - and therefore the former - will depend on the availability of limiting nutrients. Hence, in this scenario, the amount of P removed through the burial of organic sediments would be commensurate with the amount of P being supplied (which is consumed in uptake by organisms), where the latter is given by (\ref{NPA}).\\ 
{\bf Case II:} Instead, we could suppose that the organic burial flux is proportional to the rate of sedimentation per unit area, based on Figure 9.11 of \citet{SB13}. If we further consider the idealized limit wherein sedimentation can be described by Stokes' Law \citep{GM11}, the settling velocity $U$ of the particle is given by
    \begin{equation}\label{sedv}
        U = \frac{\Delta \rho a^2 g}{18 \mu_f},
    \end{equation}
where $\Delta \rho$ is the density difference between the particle and the fluid, $a$ is the diameter of the particle, and $\mu_f$ is the viscosity of the fluid. We cannot estimate $\Delta \rho$ and $a$ since these depend on organismal properties, and we therefore assume their characteristic values are similar to those on Earth; the value of $\mu_f$ is also taken to be approximately equal to that on Earth, since the medium (water) is the same. Thus, if the rate of sedimentation per unit area is proportional to $U$, we have $\mathcal{L}_P \propto g R^2$, where $\mathcal{L}_P$ in this context represents the burial rate of organic sediments (in yr$^{-1}$) and the factor of $R^2$ represents the surface area. We use the scaling $M \propto R^{3.3}$ for icy worlds with $R < R_\oplus$ \citep{SGM07} to obtain the final relation $\mathcal{L}_P \propto R^{3.3}$. Using these scalings, the depletion of P via organic burial is given by
\begin{equation}\label{LPLS}
\mathcal{L}_P \sim 3.1 \times 10^{-5}\,\mathrm{yr}^{-1}\,\left(\frac{R}{R_\oplus}\right)^{3.3},
\end{equation}
where the normalization factor is chosen such that the sink term $\mathcal{L}_P\, \mathcal{C}_P$ (with units of mol/yr) for Earth is consistent with the empirical lower bound on the depletion of P through the sedimentation of organic material \citep{PM07}. While the magnitude of (\ref{LPLS}) is comparable to (\ref{LPH}) for Earth-sized worlds with oceans that are a few kms deep, it should be noted that the scalings are different. However, as we have already remarked, because of this sink being biotic, there are even more uncertainties involved compared to abiotic sinks and sources. Although we will not undertake further analysis of this sink, it should be noted here that (\ref{LPLS}) becomes smaller than (\ref{LPH}) by about an order of magnitude for Enceladus, while the converse is true for Europa. Hence, our value of $\phi_P$ calculated (from their sum) for Europa might have to be lowered by an order of magnitude when this sink is included, but the corresponding estimate for Enceladus would be relatively unchanged.

In applying the above formula, an important criterion was assumed to be valid: the (downward) settling velocity must exceed the (upward) vertical velocity $w$ in the deep ocean. Quantitatively, this relation can be expressed as follows: 
\begin{equation} \label{Rcrit}
    \left(\frac{\Delta \rho}{200\,\mathrm{kg/m^3}}\right)\left(\frac{a}{10^{-5}\,\mathrm{m}}\right)^2 \left(\frac{R}{R_\oplus}\right)^{1.3} > 0.01 \left(\frac{w}{10^{-7}\,\mathrm{m/s}}\right).
\end{equation}
The LHS has been normalized based on the characteristic values for phytoplankton \citep{Mag13}, whereas the denominator on the RHS corresponds to the average deep ocean vertical velocity on Earth \citep{LSW17}. Hence, from (\ref{Rcrit}), it is conceivable that sedimentation of organic material could occur when $R/R_\oplus \gtrsim 0.03$.

\subsection{Sources and sinks for other nutrients}\label{SSecON}
We begin with a brief discussion of the sources and sinks for N. It is known that hydrothermal vents serve as sinks for P and several other trace metals and rare earth elements \citep{DAB13}. However, these systems are ostensibly not an abiotic sink for N. As a result, we are left with two phenomena: submarine weathering and burial of organic material that would function as a source and a sink for N respectively. 

To the best of our knowledge, we are not aware of studies that have been undertaken to estimate the dissolution rates for nitrate-producing minerals for different pH values. Hence, constructing the analog of (\ref{Rate}) is not easy and the issues with estimating the burial rate have already been delineated earlier. However, a potential way for calculating the latter is to assume that Case II is valid, in which case the sedimentation rate $\mathcal{L}_N$ would be the same as (\ref{LPLS}). This is because all particles of a given size would undergo sedimentation at a constant rate. On the other hand, note that $\mathcal{L}_P\, \mathcal{C}_P \neq \mathcal{L}_N\, \mathcal{C}_N$ even if $\mathcal{L}_P \approx \mathcal{L}_N$ because the concentrations of P and N will differ. On Earth, most organisms in the ocean are characterized by the Redfield ratio of N:P = 16:1. Although we cannot estimate the value of N produced due to submarine weathering, it appears likely to be lower than the corresponding value for continental weathering by rain water; if we consider the latter as approximated by the riverine input, it equals $\sim 10^{12}$ mol/yr for an Earth-sized world \citep{SMB10}. 

Next, we turn our attention to Fe. Although the major sources of dissolved Fe in the oceans are mineral dust \citep{JM15} and sediments from subaerial continental weathering, neither are likely to function on subsurface worlds. Instead, it would be necessary to consider the submarine weathering of Fe, which depends on the pH and will not be considered further herein. However, an interesting aspect of the iron cycle is that hydrothermal vents actually serve as a \emph{source} of dissolved Fe \citep{TBD10}. The total amount of Fe produced per unit time ($\mathcal{S}_\mathrm{Fe}$) depends on the amount of hydrothermal circulation $F_h$ that was argued to be proportional to the mass $M$ of the ocean world. Based on this ansatz, we obtain
\begin{equation}\label{NFeA}
\mathcal{S}_\mathrm{Fe} \sim \,9 \times 10^{8}\,\mathrm{mol/yr}\,\left(\frac{R}{R_\oplus}\right)^{3.3},
\end{equation}    
where the normalization has been chosen based on that of Earth \citep{TBD10}, although recent studies suggest that this value may be lower by about $20\%$ \citep{FBJ14}. Turning our attention to the sinks of Fe, the burial of organic sediments play a major role. As we have stated earlier, estimating this quantity is not an easy task because of its biotic nature. However, if the line of reasoning espoused in Case II is correct, the sedimentation rate becomes $\mathcal{L}_\mathrm{Fe} \approx \mathcal{L}_P$ for the reasons outlined earlier when discussing N sinks. 

In closing, a few general comments are in order. The solubility of water plays a highly important role in determining the availability of dissolved nutrients. In many instances, phosphates are relatively insoluble whereas nitrates are very soluble \citep{SB13}. The interesting case is iron, since it happens to be insoluble when oxidized and highly soluble in reduced form \citep{Dav93}; the same behavior is also evinced by manganese. Hence, on subsurface ocean worlds like Enceladus, where H$_2$ production has been documented, it seems likely that Fe would be soluble. Therefore, iron is rather unlikely to play the role of the limiting nutrient on such worlds. Future analyses of the limiting nutrient on subsurface ocean worlds will need to take the solubility of water into account.

\section{Detection of bioessential elements}\label{SecSearch}
Here, we will briefly discuss two important aspects of remotely searching for bioessential elements - based on our understanding of ``life as we know it'' - and the potential implications.

\subsection{Stellar spectroscopy for bioessential elements}
Given the importance of bioessential elements such as P in regulating ocean productivity, determining the stellar abundances of these elements could yield important information about the potential habitability of planets and moons orbiting the stars.

Several studies have already focused on determining P abundances for stars with varying values of [Fe/H] either via the near-ultraviolet P I doublet at 2135/2136 \AA\, \citep{JTF14} or through the weak P I lines in the infrared at $10500$-$10820$ \AA\, \citep{CBF11,MPC17}. Most of these studies obtained an average value of [P/Fe] of around $0.1$ for stars in the metallicity range of $-1.0 < \mathrm{[Fe/H]} < 0.2$. An interesting point worth noting from Figure 2 of \citet{JTF14} is that [P/H] appears to be roughly proportional to the metallicity [Fe/H] across several orders of magnitude. 

Heavier bioessential elements - one notable example being Mo, which is discussed further below -  are synthesized through a variety of mechanisms including the slow and rapid neutron-capture processes (the s- and r-process respectively) and the p-process. Recent evidence, based on LIGO's gravitational wave detection of GW170817 \citep{AAA17} and follow-up electromagnetic observations \citep{VGB}, implies that neutron star mergers play a dominant role in the production of r-processed elements \citep{KM17,CB17}. Since these events are spatially localized, the stellar abundances of such elements should vary considerably, and this is borne out by observations \citep{DMT17}. Studies of metal-poor stars using the Mo I 3864 and 5506  \AA\, spectral lines have revealed that the Mo/Fe ratio varies by more than two orders of magnitude, as seen from Figure 4 or Tables 4 and 5 of \citet{HAC14}. It is therefore evident that habitable exoplanets with bioessential element abundances very different from that of Earth will exist.

Thus, searching for spectral signatures of bioessential elements in stars known to host planets in the habitable zone may be viable and worth undertaking, but the following caveats are worth recognizing. First, the stellar and planetary abundances of bioessential elements will not necessarily be correlated since planets will typically have a wide range of compositions. Second, it does not always follow that the concentrations of these elements in the oceans will be proportional to their crustal abundances. For instance, only a tiny fraction ($\sim 10^{-4}$) of P on land or the ocean floor is actually present in the biosphere \citep{Mac05}. Despite these issues, if the stellar abundances of P and other bioessential elements can be determined, we suggest that this path is worth pursuing.

Looking beyond bioessential elements, identifying the abundances of long-lived radioactive nuclides is also an important endeavor. Inferring the abundances of these elements, which are also expected to vary significantly across stars \citep{Fre10}, may yield valuable information about mantle convection in the interior, the thickness of ice shells overlying putative subsurface ocean worlds and the potential mass of biospheres that can be supported by radiolysis \citep{HSS06,LL18}. For example, it should be feasible to measure the abundance of uranium via the U II spectral line at 3860 \AA\, \citep{FCN07}. In this context, a recent study by \citet{UJP15} concluded that the Sun is depleted in thorium compared to most solar-type stars, indicating that planets and moons orbiting them may have greater energy budgets.

\subsection{Molybdenum and the clouds of Venus}\label{SSecMo}
Although the surface of Venus is uninhabitable for ``life as we know it'', it may have been habitable until as recently as $0.715$ Ga \citep{WD16}. There have been many theoretical proposals in favor of aerial biospheres in the clouds of Venus \citep{Mor67,SG04,LMS18}, although significant challenges are ostensibly posed by physicochemical factors like acidity and solar radiation.  

The composition of Venus' clouds has been inferred through a combination of Earth-based observations, the Galileo orbiter and the Venus Express mission. None of these studies appear to have hitherto provided conclusive evidence for the existence of molybdenum (Mo) in the clouds \citep{MMP18}. Although the issue of whether Mo is truly necessary for ``life as we know it'' remains unsettled, the unique features of molybdenum chemistry and the presence of Mo in many essential enzymes suggest that it does constitute an essential biological element \citep{WDS02,SMR09}. On the other hand, it should be noted that some prokaryotes use tungsten in lieu of molybdenum \citep{KA96,Hil02}.

Hence, if the clouds of Venus possess insufficient concentrations of Mo - or perhaps tungsten (W), given that it serves as an effective substitute for Mo in certain enzymes - this environment may prove to be relatively uninhabitable or possess oligotrophic biospheres that are not readily detectable. Hence, we propose that future missions to Venus, in addition to searching for biosignatures, should also carry instruments for detecting bioessential trace elements such as Mo.

\section{Conclusions}\label{SecConc}
The presence of elements necessary for biological functions (in Earth-based organisms) in sufficiently high concentrations represents one of the requirements for habitability. In this paper, we focused on identifying the putative elements that regulate total biological productivity on worlds with subsurface oceans; in theory, if these bioessential elements are absent, the oceans ought to be devoid of ``life as we know it''.

We argued that the limiting nutrient for ocean productivity over long timescales is phosphorus (P), in the form of phosphates. By formulating a simple mathematical model that quantified the various biotic and abiotic sources and sinks for P, we concluded that its steady-state concentration is likely to be a few orders of magnitude lower relative to the Earth if the ocean worlds are fairly large, have alkaline oceans (with $\mathrm{pH} \gtrsim 8.0$) and hydrothermal activity.\footnote{While Enceladus ostensibly has an alkaline ocean, Europa appears to have a highly acidic ocean as per some models. Hence, insofar the availability of P is concerned, perhaps Europa should be assigned a higher priority compared to Enceladus.} In this scenario, we found that the oceans might possess oligotrophic biospheres, but ``life as we know it'' is still feasible. For worlds like Enceladus, we also noted that neither Fe nor N are likely to be the limiting nutrients with respect to to P. We also hypothesized that the potential lack of molybdenum in the clouds of Venus can limit the biological potential of any putative biospheres. Lastly, we argued that stellar spectroscopy constitutes a useful tool for gauging the chemical composition, and thus the biological potential, of exoplanets or exomoons around other stars.

Taken collectively, our analysis suggests that worlds with subsurface oceans are likely to have low global concentrations of biologically relevant nutrients. In other words, we may potentially expect global oligotrophic biospheres with relatively low biomass densities. However, there are several caveats that must be borne in mind: our methodology was reliant upon present-day values for the various sources and sinks, and based on a simplified model that did not incorporate all possible geophysical controls. Furthermore, the dynamical co-evolution of nutrient concentrations and ocean biogeochemical cycles was not taken into account, and there may exist major sources and sinks that are unique to subsurface ocean worlds (i.e. mostly absent on Earth). 

Although we predict that the biospheres on subsurface ocean worlds are likely to be oligotrophic, and might therefore lower the chances of being detected, this ought \emph{not} be construed as grounds for ruling out future missions to subsurface worlds like Europa and Enceladus. In fact, we would argue the opposite because our central hypothesis is both falsifiable and testable: if life is detected in high concentrations, it falsifies our model and if the converse is true, our model might provide an explanation as to why many subsurface ocean worlds are not likely to be abodes for complex biospheres.

\acknowledgments
We thank the referee, Chris McKay, for his insightful suggestions. ML also wishes to thank Andrew Knoll for beneficial conversations and Adina Paytan for helpful clarifications regarding certain aspects of the paper. This work was supported in part by grants from the Breakthrough Prize Foundation for the Starshot Initiative and Harvard University's Faculty of Arts and Sciences, and by the Institute for Theory and Computation (ITC) at Harvard University.

\appendix

\section{Further comments on phosphorus sinks and sources}
We will briefly outline a different scaling for the loss rate $\mathcal{L}_P$ due to the sedimentation sink, and comment on the ramifications of incorporating tidal heating in our simple model.\footnote{It must be noted that this Appendix is not included in the published version of the manuscript.} 

In Sec. \ref{SSecPho}, the expression for the sedimentation velocity $U$ as per Stokes's law was described. Now, suppose that the particle has to traverse the ocean depth $\mathcal{H}$ at this velocity. In this event, the characteristic timescale for the particle to settle at the ocean floor is close to $\mathcal{H}/U$, whereas the sedimentation rate $\zeta$ becomes the inverse of this quantity, i.e. we have $\zeta \sim U/\mathcal{H}$. Upon taking into consideration the fact that $\mathcal{L}_P$ and $\zeta$ both have the same dimensions (of inverse time), it might be reasonable to conjecture that $\mathcal{L}_P \propto \zeta$. In this scenario, we have
\begin{equation}\label{LPLSv2}
\mathcal{L}_P \sim 1.1 \times 10^{-4}\,\mathrm{yr}^{-1}\,\left(\frac{R}{R_\oplus}\right)^{1.3}\left(\frac{\mathcal{H}}{1\,\mathrm{km}}\right)^{-1},
\end{equation}
where the quantities in the sedimentation velocity $U$, given by (\ref{sedv}), are held fixed except for $g \propto R^{1.3}$. As noted previously in Sec. \ref{SSecPho}, the normalization ensures that $\mathcal{L}_P\, \mathcal{C}_P$ yields a value that is approximately equal to the depletion rate of P through sediment formation for the Earth. When we compare (\ref{LPH}) with (\ref{LPLSv2}), there are a couple of interesting points worth mentioning. Both these equations display the same scalings with respect to $R$ and $\mathcal{H}$, unlike (\ref{LPLS}) that is characterized by different power-law exponents. In addition, (\ref{LPH}) and (\ref{LPLSv2}) are comparable in magnitude, although the latter is potentially somewhat higher. 

Lastly, we seek to highlight an important point pertaining to the prior analysis in Sec. \ref{SSecPho}. In deriving (\ref{LPH}), it was assumed that the heating source was primarily radiogenic in nature. However, for the likes of Enceladus and Europa, tidal friction plays a dominant role. As a result, the effective value of $Q_{HV}$ will be enhanced, and this will also raise $\mathcal{L}_P$ on account of (\ref{HydCirc}) and (\ref{AbSink}). In the case of Enceladus, $\mathcal{L}_P$ might be increased by a factor of $\mathcal{O}(10)$ depending on the degree of tidal heating and rock porosity among other variables \citep{CTS17}. In turn, this will lower $\phi_P$ due to its inverse dependence on $\mathcal{L}_P$. Moreover, enhanced hydrothermal activity could lead to the ocean eventually becoming more alkaline depending on the pH and the circulation timescale of the ocean through the vents. As a consequence, both $\delta$ and $\mathcal{S}_\mathrm{P}$ would become smaller in magnitude because of (\ref{defdel}) and (\ref{NPA}) respectively. From (\ref{Cequ}), it is evident that $\phi_P$ will also be lowered because of the reduction in $\mathcal{S}_\mathrm{P}$.

Hence, owing to the above reasons, it seems plausible that the steady-state concentrations derived herein for subsurface ocean worlds akin to Enceladus (and Europa) may represent upper bounds.


\begin{thebibliography}{}
\expandafter\ifx\csname natexlab\endcsname\relax\def\natexlab#1{#1}\fi
\providecommand{\url}[1]{\href{#1}{#1}}

\bibitem[{{Abbott} {et~al.}(2017){Abbott}, {Abbott}, {Abbott}, {Acernese},
  {Ackley}, {Adams}, {Adams}, {Addesso}, {Adhikari}, {Adya}, \& et~al.}]{AAA17}
{Abbott}, B.~P., {Abbott}, R., {Abbott}, T.~D., {et~al.} 2017, Astrophys. J.
  Lett., 848, L13

\bibitem[{{Adcock} {et~al.}(2013){Adcock}, {Hausrath}, \& {Forster}}]{AHF13}
{Adcock}, C.~T., {Hausrath}, E.~M., \& {Forster}, P.~M. 2013, Nat. Geosci., 6,
  824

\bibitem[{{Altwegg} {et~al.}(2016){Altwegg}, {Balsiger}, {Bar-Nun},
  {Berthelier}, {Bieler}, {Bochsler}, {Briois}, {Calmonte}, {Combi}, {Cottin},
  {De Keyser}, {Dhooghe}, {Fiethe}, {Fuselier}, {Gasc}, {Gombosi}, {Hansen},
  {Haessig}, {Ja ckel}, {Kopp}, {Korth}, {Le Roy}, {Mall}, {Marty}, {Mousis},
  {Owen}, {Reme}, {Rubin}, {Semon}, {Tzou}, {Waite}, \& {Wurz}}]{AB16}
{Altwegg}, K., {Balsiger}, H., {Bar-Nun}, A., {et~al.} 2016, Science Advances,
  2, e1600285

\bibitem[{{Anbar}(2008)}]{Anbar}
{Anbar}, A.~D. 2008, Science, 322, 1481

\bibitem[{{Anbar} \& {Knoll}(2002)}]{AK02}
{Anbar}, A.~D., \& {Knoll}, A.~H. 2002, Science, 297, 1137

\bibitem[{{Benitez-Nelson}(2000)}]{BN00}
{Benitez-Nelson}, C.~R. 2000, Earth Sci. Rev., 51, 109

\bibitem[{{Boyd} \& {Ellwood}(2010)}]{BE10}
{Boyd}, P.~W., \& {Ellwood}, M.~J. 2010, Nature Geoscience, 3, 675

\bibitem[{{Caffau} {et~al.}(2011){Caffau}, {Bonifacio}, {Faraggiana}, \&
  {Steffen}}]{CBF11}
{Caffau}, E., {Bonifacio}, P., {Faraggiana}, R., \& {Steffen}, M. 2011, Astron.
  Astrophys., 532, A98

\bibitem[{{Choblet} {et~al.}(2017){Choblet}, {Tobie}, {Sotin}, {B{\v
  e}hounkov{\'a}}, {{\v C}adek}, {Postberg}, \& {Sou{\v c}ek}}]{CTS17}
{Choblet}, G., {Tobie}, G., {Sotin}, C., {et~al.} 2017, Nat. Astron., 1, 841

\bibitem[{{Chornock} {et~al.}(2017){Chornock}, {Berger}, {Kasen},
  {Cowperthwaite}, {Nicholl}, {Villar}, {Alexander}, {Blanchard}, {Eftekhari},
  {Fong}, {Margutti}, {Williams}, {Annis}, {Brout}, {Brown}, {Chen}, {Drout},
  {Farr}, {Foley}, {Frieman}, {Fryer}, {Herner}, {Holz}, {Kessler}, {Matheson},
  {Metzger}, {Quataert}, {Rest}, {Sako}, {Scolnic}, {Smith}, \&
  {Soares-Santos}}]{CB17}
{Chornock}, R., {Berger}, E., {Kasen}, D., {et~al.} 2017, Astrophys. J. Lett.,
  848, L19

\bibitem[{{Chyba} \& {Hand}(2001)}]{CH01}
{Chyba}, C.~F., \& {Hand}, K.~P. 2001, Science, 292, 2026

\bibitem[{{Cockell} {et~al.}(2016){Cockell}, {Bush}, {Bryce}, {Direito},
  {Fox-Powell}, {Harrison}, {Lammer}, {Landenmark}, {Martin-Torres},
  {Nicholson}, {Noack}, {O'Malley-James}, {Payler}, {Rushby}, {Samuels},
  {Schwendner}, {Wadsworth}, \& {Zorzano}}]{CB16}
{Cockell}, C.~S., {Bush}, T., {Bryce}, C., {et~al.} 2016, Astrobiology, 16, 89

\bibitem[{{Cox} {et~al.}(2018){Cox}, {Lyons}, {Mitchell}, {Hasterok}, \&
  {Gard}}]{CLM18}
{Cox}, G.~M., {Lyons}, T.~W., {Mitchell}, R.~N., {Hasterok}, D., \& {Gard}, M.
  2018, Earth Planet. Sci. Lett., 489, 28

\bibitem[{{Davison}(1993)}]{Dav93}
{Davison}, W. 1993, Earth Sci. Rev., 34, 119

\bibitem[{{Delaney}(1998)}]{Del98}
{Delaney}, M.~L. 1998, Global Biogeochem. Cycles, 12, 563

\bibitem[{{Delgado Mena} {et~al.}(2017){Delgado Mena}, {Tsantaki}, {Adibekyan},
  {Sousa}, {Santos}, {Gonz{\'a}lez Hern{\'a}ndez}, \& {Israelian}}]{DMT17}
{Delgado Mena}, E., {Tsantaki}, M., {Adibekyan}, V.~Z., {et~al.} 2017, Astron.
  Astrophys., 606, A94

\bibitem[{{Dick} {et~al.}(2013){Dick}, {Anantharaman}, {Baker}, {Li}, {Reed},
  \& {Sheik}}]{DAB13}
{Dick}, G.~J., {Anantharaman}, K., {Baker}, B.~J., {et~al.} 2013, Front.
  Microbiol., 4, 124

\bibitem[{{Fitzsimmons} {et~al.}(2014){Fitzsimmons}, {Boyle}, \&
  {Jenkins}}]{FBJ14}
{Fitzsimmons}, J.~N., {Boyle}, E.~A., \& {Jenkins}, W.~J. 2014, Proc. Natl.
  Acad. Sci. USA, 111, 16654

\bibitem[{{Frebel}(2010)}]{Fre10}
{Frebel}, A. 2010, Astron. Nachr., 331, 474

\bibitem[{{Frebel} {et~al.}(2007){Frebel}, {Christlieb}, {Norris}, {Thom},
  {Beers}, \& {Rhee}}]{FCN07}
{Frebel}, A., {Christlieb}, N., {Norris}, J.~E., {et~al.} 2007, Astrophys. J.
  Lett., 660, L117

\bibitem[{{Froelich} {et~al.}(1982){Froelich}, {Bender}, {Luedtke}, {Heath}, \&
  {DeVries}}]{FBL82}
{Froelich}, P.~N., {Bender}, M.~L., {Luedtke}, N.~A., {Heath}, G.~R., \&
  {DeVries}, T. 1982, Am. J. Sci., 282, 474

\bibitem[{{Glein} {et~al.}(2015){Glein}, {Baross}, \& {Waite}}]{GBW15}
{Glein}, C.~R., {Baross}, J.~A., \& {Waite}, J.~H. 2015, Geochim. Cosmochim.
  Acta, 162, 202

\bibitem[{{Guazzelli} \& {Morris}(2011)}]{GM11}
{Guazzelli}, E., \& {Morris}, J.~F. 2011, Cambridge Texts in Applied
  Mathematics, Vol.~45, {A Physical Introduction to Suspension Dynamics}
  (Cambridge Univ. Press)

\bibitem[{{Hansen} {et~al.}(2014){Hansen}, {Andersen}, \& {Christlieb}}]{HAC14}
{Hansen}, C.~J., {Andersen}, A.~C., \& {Christlieb}, N. 2014, Astron.
  Astrophys., 568, A47

\bibitem[{{Hille}(2002)}]{Hil02}
{Hille}, R. 2002, Trends Biochem. Sci., 27, 360

\bibitem[{{Hoehler} {et~al.}(2007){Hoehler}, {Amend}, \& {Shock}}]{HAS07}
{Hoehler}, T.~M., {Amend}, J.~P., \& {Shock}, E.~L. 2007, Astrobiology, 7, 819

\bibitem[{{Hsu} {et~al.}(2015){Hsu}, {Postberg}, {Sekine}, {Shibuya}, {Kempf},
  {Hor{\'a}nyi}, {Juh{\'a}sz}, {Altobelli}, {Suzuki}, {Masaki}, {Kuwatani},
  {Tachibana}, {Sirono}, {Moragas-Klostermeyer}, \& {Srama}}]{HP15}
{Hsu}, H.-W., {Postberg}, F., {Sekine}, Y., {et~al.} 2015, Nature, 519, 207

\bibitem[{{Hudson} {et~al.}(2000){Hudson}, {Taylor}, \& {Schindler}}]{HTS00}
{Hudson}, J.~J., {Taylor}, W.~D., \& {Schindler}, D.~W. 2000, Nature, 406, 54

\bibitem[{{Hussmann} {et~al.}(2006){Hussmann}, {Sohl}, \& {Spohn}}]{HSS06}
{Hussmann}, H., {Sohl}, F., \& {Spohn}, T. 2006, Icarus, 185, 258

\bibitem[{{Jacobson} {et~al.}(2014){Jacobson}, {Thanathibodee}, {Frebel},
  {Roederer}, {Cescutti}, \& {Matteucci}}]{JTF14}
{Jacobson}, H.~R., {Thanathibodee}, T., {Frebel}, A., {et~al.} 2014, Astrophys.
  J. Lett., 796, L24

\bibitem[{{Jickells} \& {Moore}(2015)}]{JM15}
{Jickells}, T., \& {Moore}, C.~M. 2015, Annu. Rev. Ecol. Evol. Syst., 46, 481

\bibitem[{{Karl} \& {Bj{\"o}rkman}(2015)}]{KB14}
{Karl}, D.~M., \& {Bj{\"o}rkman}, K.~M. 2015, in {Biogeochemistry of Marine
  Dissolved Organic Matter}, 2nd edn., ed. D.~A. {Hansell} \& C.~A. {Carlson}
  (Elsevier), 233--334

\bibitem[{{Kasen} {et~al.}(2017){Kasen}, {Metzger}, {Barnes}, {Quataert}, \&
  {Ramirez-Ruiz}}]{KM17}
{Kasen}, D., {Metzger}, B., {Barnes}, J., {Quataert}, E., \& {Ramirez-Ruiz}, E.
  2017, Nature, 551, 80

\bibitem[{{Kipp} \& {St{\"u}eken}(2017)}]{KS17}
{Kipp}, M.~A., \& {St{\"u}eken}, E.~E. 2017, Sci. Adv., 3, eaao4795

\bibitem[{{Kletzin} \& {Adams}(1996)}]{KA96}
{Kletzin}, A., \& {Adams}, M.~W.~W. 1996, FEMS Microbiol. Rev., 18, 5

\bibitem[{{Knoll}(2017)}]{Kno17}
{Knoll}, A.~H. 2017, Nature, 548, 528

\bibitem[{{Liang} {et~al.}(2017){Liang}, {Spall}, \& {Wunsch}}]{LSW17}
{Liang}, X., {Spall}, M., \& {Wunsch}, C. 2017, J. Geophys. Res. C, 122, 8208

\bibitem[{{Limaye} {et~al.}(2018){Limaye}, {Mogul}, {Smith}, {Ansari},
  {S{\l}owik}, \& {Vaishampayan}}]{LMS18}
{Limaye}, S.~S., {Mogul}, R., {Smith}, D.~J., {et~al.} 2018, Astrobiology,
  doi:10.1089/ast.2017.1783

\bibitem[{{Lingam} \& {Loeb}(2018)}]{LL18}
{Lingam}, M., \& {Loeb}, A. 2018, Int. J. Astrobiol., doi:10.1017/
  S1473550418000083

\bibitem[{{Lunine}(2017)}]{Lun17}
{Lunine}, J.~I. 2017, Acta Astronaut., 131, 123

\bibitem[{{Maas} {et~al.}(2017){Maas}, {Pilachowski}, \& {Cescutti}}]{MPC17}
{Maas}, Z.~G., {Pilachowski}, C.~A., \& {Cescutti}, G. 2017, Astrophys. J.,
  841, 108

\bibitem[{{Maci{\'a}}(2005)}]{Mac05}
{Maci{\'a}}, E. 2005, Chem. Soc. Rev., 34, 691

\bibitem[{{Maggi}(2013)}]{Mag13}
{Maggi}, F. 2013, J. Geophys. Res. C, 118, 2118

\bibitem[{{Marcq} {et~al.}(2018){Marcq}, {Mills}, {Parkinson}, \&
  {Vandaele}}]{MMP18}
{Marcq}, E., {Mills}, F.~P., {Parkinson}, C.~D., \& {Vandaele}, A.~C. 2018,
  Space Sci. Rev., 214, \#10

\bibitem[{{Marion} {et~al.}(2003){Marion}, {Fritsen}, {Eicken}, \&
  {Payne}}]{MF03}
{Marion}, G.~M., {Fritsen}, C.~H., {Eicken}, H., \& {Payne}, M.~C. 2003,
  Astrobiology, 3, 785

\bibitem[{{Martin} {et~al.}(2008){Martin}, {Baross}, {Kelley}, \&
  {Russell}}]{MBK08}
{Martin}, W., {Baross}, J., {Kelley}, D., \& {Russell}, M.~J. 2008, Nat. Rev.
  Microbiol., 6, 805

\bibitem[{{McKay} {et~al.}(2008){McKay}, {Porco}, {Altheide}, {Davis}, \&
  {Kral}}]{MPA08}
{McKay}, C.~P., {Porco}, C.~C., {Altheide}, T., {Davis}, W.~L., \& {Kral},
  T.~A. 2008, Astrobiology, 8, 909

\bibitem[{{Moore} {et~al.}(2013){Moore}, {Mills}, {Arrigo}, {Berman-Frank},
  {Bopp}, {Boyd}, {Galbraith}, {Geider}, {Guieu}, {Jaccard}, {Jickells}, {La
  Roche}, {Lenton}, {Mahowald}, {Mara{\~n}{\'o}n}, {Marinov}, {Moore},
  {Nakatsuka}, {Oschlies}, {Saito}, {Thingstad}, {Tsuda}, \& {Ulloa}}]{MMA13}
{Moore}, C.~M., {Mills}, M.~M., {Arrigo}, K.~R., {et~al.} 2013, Nat. Geosci.,
  6, 701

\bibitem[{{Morowitz} \& {Sagan}(1967)}]{Mor67}
{Morowitz}, H., \& {Sagan}, C. 1967, Nature, 215, 1259

\bibitem[{{Moutin} {et~al.}(2002){Moutin}, {Thingstad}, {Van Wambeke}, {Marie},
  {Slawyk}, {Raimbault}, \& {Claustre}}]{MT02}
{Moutin}, T., {Thingstad}, T.~F., {Van Wambeke}, F., {et~al.} 2002, Limnol.
  Oceanogr., 47, 1562

\bibitem[{{Nimmo} \& {Pappalardo}(2016)}]{NP16}
{Nimmo}, F., \& {Pappalardo}, R.~T. 2016, J. Geophys. Res. E, 121, 1378

\bibitem[{{Pasek} \& {Greenberg}(2012)}]{PG12}
{Pasek}, M.~A., \& {Greenberg}, R. 2012, Astrobiology, 12, 151

\bibitem[{{Pasek} {et~al.}(2017){Pasek}, {Gull}, \& {Herschy}}]{PGH17}
{Pasek}, M.~A., {Gull}, M., \& {Herschy}, B. 2017, Chem. Geol., 475, 149

\bibitem[{{Patey} {et~al.}(2008){Patey}, {Rijkenberg}, {Statham},
  {Stinchcombe}, {Achterberg}, \& {Mowlem}}]{PRS08}
{Patey}, M.~D., {Rijkenberg}, M.~J.~A., {Statham}, P.~J., {et~al.} 2008, Trends
  Anal. Chem., 27, 169

\bibitem[{{Paytan} \& {McLaughlin}(2007)}]{PM07}
{Paytan}, A., \& {McLaughlin}, K. 2007, Chem. Rev., 107, 563

\bibitem[{{Planavsky} {et~al.}(2010){Planavsky}, {Rouxel}, {Bekker}, {Lalonde},
  {Konhauser}, {Reinhard}, \& {Lyons}}]{PRB10}
{Planavsky}, N.~J., {Rouxel}, O.~J., {Bekker}, A., {et~al.} 2010, Nature, 467,
  1088

\bibitem[{{Porco} {et~al.}(2017){Porco}, {Dones}, \& {Mitchell}}]{PDM17}
{Porco}, C.~C., {Dones}, L., \& {Mitchell}, C. 2017, Astrobiology, 17, 876

\bibitem[{{Postberg} {et~al.}(2018){Postberg}, {Khawaja}, {Abel}, {Choblet},
  {Glein}, {Gudipati}, {Henderson}, {Hsu}, {Kempf}, {Klenner},
  {Moragas-Klostermeyer}, {Magee}, {N{\"a}lle}, {Perry}, {Reviol}, {Schmidt},
  {Srama}, {Stolz}, {Tobie}, {Trieloff}, \& {Waite}}]{PKA18}
{Postberg}, F., {Khawaja}, N., {Abel}, B., {et~al.} 2018, Nature, 558, 564

\bibitem[{{Reinhard} {et~al.}(2017){Reinhard}, {Planavsky}, {Gill}, {Ozaki},
  {Robbins}, {Lyons}, {Fischer}, {Wang}, {Cole}, \& {Konhauser}}]{RPG17}
{Reinhard}, C.~T., {Planavsky}, N.~J., {Gill}, B.~C., {et~al.} 2017, Nature,
  541, 386

\bibitem[{{Robbins} {et~al.}(2016){Robbins}, {Lalonde}, {Planavsky}, {Partin},
  {Reinhard}, {Kendall}, {Scott}, {Hardisty}, {Gill}, {Alessi}, {Dupont},
  {Saito}, {Crowe}, {Poulton}, {Bekker}, {Lyons}, \& {Konhauser}}]{RLP16}
{Robbins}, L.~J., {Lalonde}, S.~V., {Planavsky}, N.~J., {et~al.} 2016,
  Earth-Sci. Rev., 163, 323

\bibitem[{{Russell} {et~al.}(2014){Russell}, {Barge}, {Bhartia}, {Bocanegra},
  {Bracher}, {Branscomb}, {Kidd}, {McGlynn}, {Meier}, {Nitschke}, {Shibuya},
  {Vance}, {White}, \& {Kanik}}]{RBB14}
{Russell}, M.~J., {Barge}, L.~M., {Bhartia}, R., {et~al.} 2014, Astrobiology,
  14, 308

\bibitem[{{Sarmiento} \& {Gruber}(2006)}]{SG06}
{Sarmiento}, J.~L., \& {Gruber}, N. 2006, {Ocean Biogeochemical Dynamics}
  (Princeton Univ. Press)

\bibitem[{{Schlesinger} \& {Bernhardt}(2013)}]{SB13}
{Schlesinger}, W.~H., \& {Bernhardt}, E.~S. 2013, {Biogeochemistry: An Analysis
  of Global Change} (Academic Press)

\bibitem[{{Schulze-Makuch} {et~al.}(2004){Schulze-Makuch}, {Grinspoon},
  {Abbas}, {Irwin}, \& {Bullock}}]{SG04}
{Schulze-Makuch}, D., {Grinspoon}, D.~H., {Abbas}, O., {Irwin}, L.~N., \&
  {Bullock}, M.~A. 2004, Astrobiology, 4, 11

\bibitem[{{Schulze-Makuch} \& {Irwin}(2008)}]{SI08}
{Schulze-Makuch}, D., \& {Irwin}, L.~N. 2008, {Life in the Universe:
  Expectations and Constraints} (Springer-Verlag),
  doi:10.1007/978-3-540-76817-3

\bibitem[{{Schwarz} {et~al.}(2009){Schwarz}, {Mendel}, \& {Ribbe}}]{SMR09}
{Schwarz}, G., {Mendel}, R.~R., \& {Ribbe}, M.~W. 2009, Nature, 460, 839

\bibitem[{{Seitzinger} {et~al.}(2010){Seitzinger}, {Mayorga}, {Bouwman},
  {Kroeze}, {Beusen}, {Billen}, {van Drecht}, {Dumont}, {Fekete}, {Garnier}, \&
  {Harrison}}]{SMB10}
{Seitzinger}, S.~P., {Mayorga}, E., {Bouwman}, A.~F., {et~al.} 2010, Global
  Biogeochem. Cycles, 24, GB0A08

\bibitem[{{Sojo} {et~al.}(2016){Sojo}, {Herschy}, {Whicher}, {Camprub{\'{\i}}},
  \& {Lane}}]{SHW16}
{Sojo}, V., {Herschy}, B., {Whicher}, A., {Camprub{\'{\i}}}, E., \& {Lane}, N.
  2016, Astrobiology, 16, 181

\bibitem[{{Sotin} {et~al.}(2007){Sotin}, {Grasset}, \& {Mocquet}}]{SGM07}
{Sotin}, C., {Grasset}, O., \& {Mocquet}, A. 2007, Icarus, 191, 337

\bibitem[{{Sparks} {et~al.}(2017){Sparks}, {Schmidt}, {McGrath}, {Hand},
  {Spencer}, {Cracraft}, \& {E Deustua}}]{SS17}
{Sparks}, W.~B., {Schmidt}, B.~E., {McGrath}, M.~A., {et~al.} 2017, Astrophys.
  J. Lett., 839, L18

\bibitem[{{Steel} {et~al.}(2017){Steel}, {Davila}, \& {McKay}}]{SDM17}
{Steel}, E.~L., {Davila}, A., \& {McKay}, C.~P. 2017, Astrobiology, 17, 862

\bibitem[{{Stein} \& {Stein}(1994)}]{SS94}
{Stein}, C.~A., \& {Stein}, S. 1994, J. Geophys. Res., 99, 3081

\bibitem[{{Tagliabue} {et~al.}(2017){Tagliabue}, {Bowie}, {Boyd}, {Buck},
  {Johnson}, \& {Saito}}]{TBB17}
{Tagliabue}, A., {Bowie}, A.~R., {Boyd}, P.~W., {et~al.} 2017, Nature, 543, 51

\bibitem[{{Tagliabue} {et~al.}(2010){Tagliabue}, {Bopp}, {Dutay}, {Bowie},
  {Chever}, {Jean-Baptiste}, {Bucciarelli}, {Lannuzel}, {Remenyi}, {Sarthou},
  {Aumont}, {Gehlen}, \& {Jeandel}}]{TBD10}
{Tagliabue}, A., {Bopp}, L., {Dutay}, J.-C., {et~al.} 2010, Nat. Geosci., 3,
  252

\bibitem[{{Tyrrell}(1999)}]{Tyrr99}
{Tyrrell}, T. 1999, Nature, 400, 525

\bibitem[{{Unterborn} {et~al.}(2015){Unterborn}, {Johnson}, \&
  {Panero}}]{UJP15}
{Unterborn}, C.~T., {Johnson}, J.~A., \& {Panero}, W.~R. 2015, Astrophys. J.,
  806, 139

\bibitem[{{Valencia} \& {O'Connell}(2009)}]{VOC09}
{Valencia}, D., \& {O'Connell}, R.~J. 2009, Earth Planet. Sci. Lett., 286, 492

\bibitem[{{Villar} {et~al.}(2017){Villar}, {Guillochon}, {Berger}, {Metzger},
  {Cowperthwaite}, {Nicholl}, {Alexander}, {Blanchard}, {Chornock},
  {Eftekhari}, {Fong}, {Margutti}, \& {Williams}}]{VGB}
{Villar}, V.~A., {Guillochon}, J., {Berger}, E., {et~al.} 2017, Astrophys. J.
  Lett., 851, L21

\bibitem[{{Waite} {et~al.}(2017){Waite}, {Glein}, {Perryman}, {Teolis},
  {Magee}, {Miller}, {Grimes}, {Perry}, {Miller}, {Bouquet}, {Lunine},
  {Brockwell}, \& {Bolton}}]{WG17}
{Waite}, J.~H., {Glein}, C.~R., {Perryman}, R.~S., {et~al.} 2017, Science, 356,
  155

\bibitem[{{Way} {et~al.}(2016){Way}, {Del Genio}, {Kiang}, {Sohl}, {Grinspoon},
  {Aleinov}, {Kelley}, \& {Clune}}]{WD16}
{Way}, M.~J., {Del Genio}, A.~D., {Kiang}, N.~Y., {et~al.} 2016, Geophys. Res.
  Lett., 43, 8376

\bibitem[{{Westall} \& {Brack}(2018)}]{WB18}
{Westall}, F., \& {Brack}, A. 2018, Space Sci. Rev., 214, 50

\bibitem[{{Wheat} {et~al.}(1996){Wheat}, {Feely}, \& {Mottl}}]{WFM96}
{Wheat}, C.~G., {Feely}, R.~A., \& {Mottl}, M.~J. 1996, Geochim. Cosmochim.
  Acta, 60, 3593

\bibitem[{{Wheat} {et~al.}(2003){Wheat}, {McManus}, {Mottl}, \&
  {Giambalvo}}]{WM03}
{Wheat}, C.~G., {McManus}, J., {Mottl}, M.~J., \& {Giambalvo}, E. 2003,
  Geophys. Res. Lett., 30, 1895

\bibitem[{{Williams} \& {Frausto Da Silva}(2002)}]{WDS02}
{Williams}, R.~J.~P., \& {Frausto Da Silva}, J.~J.~R. 2002, Biochem. Biophys.
  Res. Commun., 292, 293

\bibitem[{{Zhang} \& {Chi}(2002)}]{ZC02}
{Zhang}, J.-Z., \& {Chi}, J. 2002, Environ. Sci. Technol., 36, 1048

\end{thebibliography}

\end{document}